# An Integrated Magnetics Design for an Isolated ZVS Ćuk Converter

Sadegh Esmaeili Rad, *Graduate Student Member, IEEE*, and Sudip K Mazumder, *Fellow, IEEE*

***Abstract*—This paper proposes a new integrated magnetics (IM) design for an isolated zero-voltage-switching (ZVS) Ćuk converter (IZCC). In this design, six magnets are wound onto a single magnetic core, and to minimize magnetic core size and losses, both direct current (DC) and alternating current (AC) flux cancellations are considered. The DC flux is fully cancelled, and the AC flux must be cancelled until a limited value such that the input and output inductor currents have enough ripple to provide the conditions for achieving ZVS on all switches. Therefore, the value of the coupling coefficients (CC) between the windings should be considered such that the minimum ripple to achieve ZVS for all the switches is available. The design is implemented on a simple magnetic U-core, and the CC values are specified based on the winding locations and arrangement. To validate the idea experimentally, a hardware prototype is proposed with a power of 0.5 kW, a switching frequency of 150 *k*Hz, and a peak efficiency of 97.25%.**

## I. Introduction

Passive components play a critical role in power conversion systems [1][2]. These components primarily include magnets and capacitors. Magnetics are components that provide energy savings, isolation, voltage conversion, and filtering. Depending on the type of power electronics converter, the number of these magnetics could change. Therefore, the way of designing and implementing these components has a direct effect on the converter operation and performance [3][4].

Besides their importance in power conversion, magnetics also significantly affect converter volume and efficiency. The magnetics are usually bulky components that need to be designed properly [5]. There are different approaches to efficiently and compactly designing magnetic components in power converters. The conventional approach is a discrete approach, in which each magnetic is designed with a separate magnetic core. Although this approach is simple, it requires more numbers of cores, and the cores need more volume, which affects the power density [6]. Besides that, adding more magnetic cores increases the cost of components and also, and each core has its own core losses, which increases the overall losses [7].

To overcome the mentioned problems of the discrete approach in magnetic designs, the approach of integration of magnetics in a single core has been proposed [8]. Although this approach increases the complexity in design, it helps in reducing the number of utilized cores and increasing the power density, and it can also be effective in reducing the losses as well [9][10][11]. There are different approaches to integrating magnetics into a converter. The integration of the resonant inductor with the transformer in the resonant converters is one of the common approaches for increasing the power density and reducing component count and cost [12][13]. However, this approach may come at the expense of increased core loss of the transformer and deteriorate the performance of the power conversion system [14][15]. The references [16][17][18] shows the integration of inductors in interleaved boost converters with directly coupled inductors (CI). However, the direct-coupled inductors approach reduces component count and cost, but the core tends to saturate, requiring a larger core, especially in high-power applications.

Flux cancellation in the IM design is the approach to reduce the losses and increase the power density [19][20][21]. The flux cancellation includes two types: DC and AC flux cancellation. The DC flux in the magnetic core is effective in increasing the core volume, and therefore, DC flux cancellation helps in reducing the core volume [22][23][24]. However, in these references, DC flux cancellation approaches have increased the AC current ripple, thereby increasing core losses. The AC flux is not only effective in the core volume, but it is also the cause of the core losses because the core losses are related to the flux density variations [25]. [26], [27] show the AC flux cancellation approach. In [27], it is shown that the current ripple in the input and output inductors can be eliminated by integrating the inductors with the transformer windings. Although this approach reduces core losses, the DC flux remains, preventing a reduction in core volume.

There are different types of IM designs to reach both AC and DC flux cancellation on a specific E core in [28] that both inductors are wound on the center limb. However, the leakage inductor in this approach is very small, and an auxiliary inductor is needed to create more ripple. In order to adjust the current ripple and reach both AC and DC flux cancellation, the reference [29] has proposed the IM design on a U core. In this approach, DC flux cancellation can be easily achieved with a negative CC, and the amount of AC flux cancellation can be adjusted based on the winding arrangement and location. However, this approach is applicable when the inductor current waveforms are aligned and vary together.

Reference [27] narrated an IM approach for an isolated Ćuk converter. This approach uses the transformer windings solely for AC flux cancellation, with operation under hard-switching conditions. As shown, the CI itself can cancel current ripples at a certain value, and to achieve zero ripple in the inductor currents, the input and output inductor windings need to be integrated with the transformer windings. To make the Ćuk converter operate in a soft-switching condition, the reference

The authors are with the Department of Electrical and Computer Engineering, University of Illinois Chicago (UIC), Chicago/Illinois 60607 USA (e-mail: sesmai2@uic.edu; mazumder@uic.edu ).

[30] proposed an IM design approach with ZVS achievement, but this approach is not applicable to high-power applications due to increased circulating current, and also, the proposed IM structure helps only with AC flux cancellation. The other approach to making ZVS Ćuk is the proposed IZCC topology shown in Fig. 1 and reference [31]. By adding auxiliary circuits and a resonant inductor to the isolated Ćuk converter, the modes of operation will change, with two additional modes emerging. Therefore, the way magnetics are integrated for AC flux cancellation across all modes of operation needs to change. The approach proposed in [32] for the isolated resonant Ćuk converter utilizes both winding and core sharing to achieve DC and AC flux cancellations. In this approach, the resonant inductor is not integrated with the other magnetics, and the AC flux cannot be cancelled in phase-shift modes. Therefore, the current ripples are not very controllable and cannot be completely cancelled.

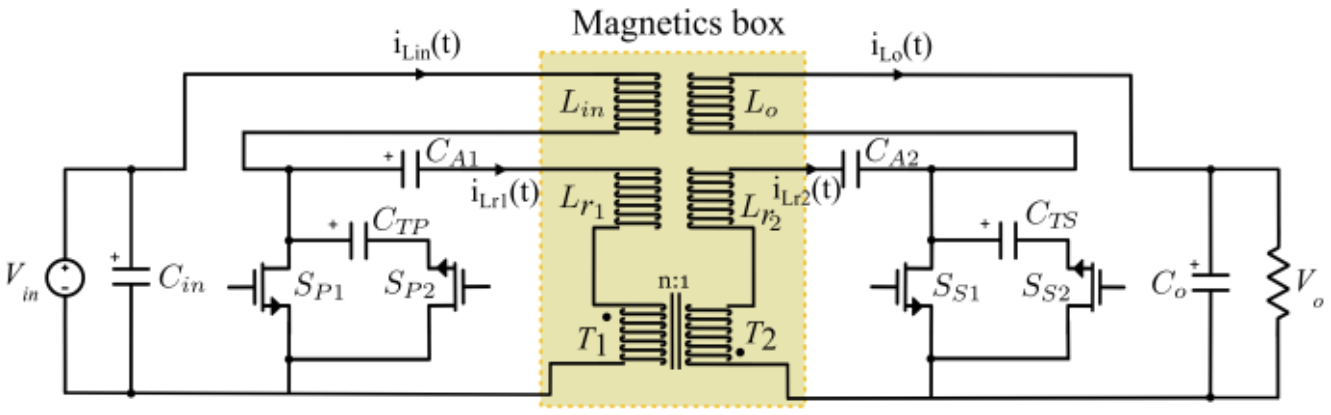

Fig. 1. IZCC topology with the magnetics box

To address the aforementioned problems, this paper proposes an IM design that integrates all six windings into a single core and uses resonant inductors (RIs) to cancel ripple across all operating modes. All the windings are wound on a single U-core. Adjusting the proper CCs between the windings is critical to have the desired amount of ripple cancellation and proper operation of the IZCC. As mentioned in [31] and [33], the input and output inductor currents need to have a minimum ripple to achieve ZVS on all switches. This is crucial when the converter operates at a high switching frequency. Therefore, the winding arrangement on the core should be such that not only is DC flux cancellation achievable, but also the CCs are adjusted to minimize ripple in the input and output inductors' current, thereby enabling ZVS. Thus, all the magnetics are integrated into a single core, and DC and AC flux cancellation is also achievable. These conditions reduce core volume, losses, costs, and the number of components.

To our knowledge, there is no prior work on integrating the different types of magnetics into a single, simple U-core that achieves both DC and AC flux cancellation and ensures ZVS for all switches. As such, the main contributions of the study are as follows:

1) Integrating all six magnetic windings of the IZCC on a single U core and winding arrangement such that both DC and AC flux cancellation happen in the magnetic core, besides proper operation of the converter;

2) Achieving ZVS on all the switches by adjusting the CCs such that the minimum current ripples are available for satisfying the ZVS conditions; and

3) 43% reduction in core volume, besides improvement in efficiency in the designed IM compared to the IZCC operation with discrete magnetics design.

## II. IM FOR SOFT SWITCHING

The AC flux cancellation in the IM design should be such that neither affects the overall operation of the converter nor increases converter losses, such as switching losses, especially when the converter operates at a high switching frequency. Therefore, the proposed approach in [27] needs to meet the soft-switching requirements. The IZCC topology is the ZVS version of the isolated Ćuk converter. This converter has four modes of operation, as shown in Fig. 2 the input and output inductor currents are not aligned at modes 1 and 3. Also, as shown in the transformer voltage waveforms in this figure, the transformer is zero in these two modes of operation. Additionally, when the DC flux of the CI is cancelled inside the core, the AC flux direction changes from positive to negative, and the transformer windings will not be able to cancel the inductors' fluxes when their flux directions change because the transformer windings' fluxes are always in one direction and only their values change. Therefore, integrating transformers with input and output inductors for AC ripple cancellation is not applicable when DC flux cancellation is desired. Thus, another winding arrangement and integration approach is needed.

This paper proposes a new integration of magnetics in the IZCC to achieve not only DC flux cancellation but also AC flux cancellation across all modes of operation. In this approach, the input and output inductors are coupled to form CI, and the RIs are also integrated on the core to help cancel AC flux. The CI will perform DC flux cancellation and AC flux cancellation in two operational modes (2 and 4). The RIs are added to cancel the AC flux in the other two modes of operation that the CI cannot perform (modes 1 and 3). The transformer is also integrated with the other magnetics with a high coupling between the primary and secondary windings. It reduces the magnetizing flux to a level that does not affect the operation of the CI and RIs. The transformer is integrated on the same core to not only reduce the overall magnetics volume but also help stabilize converter operation.

This approach offers flexibility, allowing the coupling between any two windings to be adjusted to control ripple cancellation. As the IZCC will operate at a high switching frequency, it is essential to achieve soft switching across all switches to achieve high efficiency. As the ZVS conditions rely on the input and output inductor current ripples in IZCC, there should be a tradeoff between the amount of current ripple cancellation and the ability to achieve ZVS on all switches. The current ripple on the CI currents increases the core losses. Therefore, a comparison of core and switching losses at different switching frequencies is needed to determine which is dominant and decide whether to prioritize core loss reduction or achieving ZVS on the switches.

## III. IZCC OPERATION AND ZVS CONDITIONS

### A. Operation

As shown in

Fig. 1, the IZCC converter is drawn with a magnetics box. This box has six magnetic windings, which are input ($L_{in}$) and output ($L_o$) inductors, primary ($L_{r1}$) and secondary ($L_{r2}$) side resonant inductors, primary ($T_1$) and secondary ($T_2$) windings of the transformer. These magnetics can be in different

integrated or discrete combinations. The magnetics box should be designed so that it does not affect the converter's operation.

Fig. 2 shows the IZCC operation waveforms considering all discrete magnetics. The secondary side switches ($S_{s1}, S_{s2}$) do the switching by a phase shift delay ($\rho T_{sw}$) after the primary side switches ($S_{p1}, S_{p2}$) turned off respectively. As can be seen in this figure, $\rho$ is the phase shift ratio and $T_{sw}$ is the switching period. There are two duty cycle ratios for the primary ($d_p$) and secondary ($d_s$) side switches. The switching sequence for all types of discrete or integrated magnetics is the same as Fig. 2. In [31], all modes of operation are proposed for discrete magnetics. In here, four of the main modes of operation are considered, as the deadtime modes can be ignored compared to the main modes.

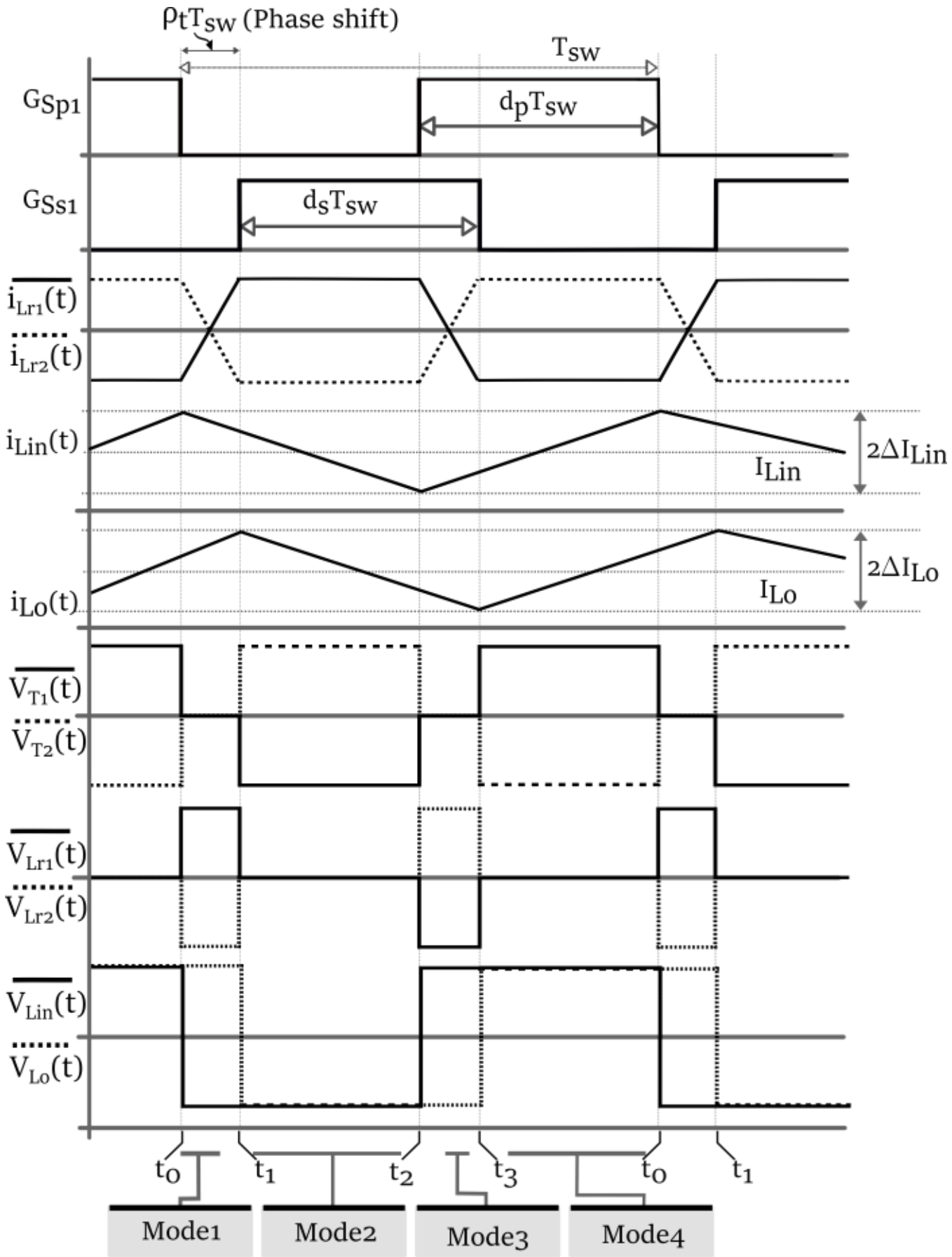

Fig. 2. IZCC waveforms of operation considering discrete magnetics

### B. ZVS Conditions

In [31], [33] a complete model of the IZCC converter is proposed, and based on the model, the ZVS conditions for the switches in IZCC are governed by (1)-(4). In these conditions, $C_{oss}$ is the switch's output capacitor, $V_{C_{TP}}$ is the average value of the primary side of the auxiliary capacitor. $V_{C_{TS}}$ is the average value of the secondary side of the auxiliary capacitor. $t_{db}$ is the deadtime between each complementary switch. $I_{Lin}$ is the average value of the input inductor current. $I_{Lo}$ is the average value of the output inductor current. $\Delta I_{Lin}$ is the ripple value of the input inductor current. $\Delta I_{Lo}$ is the ripple value of the output inductor current [31]. Parameters $i_{Lr}(t_0)$ to $i_{Lr}(t_3)$ are the initial values of the resonant inductor current at each mode of operation. It is assumed that $L_{r2}$ is transferred to the primary side and is combined with $L_{r1}$ and them $L_r = L_{r1} + n^2 L_{r2}$, where, $n$ is the transformer turns ratio.

$$i_{Lr}(t_0) < (I_{Lin} + \Delta I_{Lin}) - \frac{2C_{oss}V_{C_{TP}}}{t_{db}} \quad \text{[ZVS for } S_{P2}] \tag{1}$$

$$i_{Lr}(t_1) > -(I_{Lo} + \Delta I_{Lo}) + \frac{2C_{oss}V_{C_{TS}}}{t_{db}} \quad \text{[ZVS for } S_{S1}] \tag{2}$$

$$i_{Lr}(t_2) > (I_{Lin} - \Delta I_{Lin}) + \frac{2C_{oss}V_{C_{TP}}}{t_{db}} \quad \text{[ZVS for } S_{P1}] \tag{3}$$

$$i_{Lr}(t_3) < -(I_{Lo} - \Delta I_{Lo}) - \frac{2C_{oss}V_{C_{TS}}}{t_{db}} \quad \text{[ZVS for } S_{S2}] \tag{4}$$

The initial values of $i_{Lr}(t_0)$-$i_{Lr}(t_3)$ are specified as (5) to (8). In these equations, $d'_p = 1 - d_p$ and $d'_s = 1 - d_s$, and $I_{Lin}$ and $I_{Lo}$ are the average values of the input and output inductor currents.

$$i_{Lr}(t_0) = \frac{\begin{pmatrix} -0.5d_p d'_p V_{C_{TP}} - 0.5d_s d'_s V_{C_{TS}} \\ +(d'_s - \rho)d_{s1}V_{C_{TS}} \end{pmatrix}}{f_{sw}L_r} \tag{5}$$

$$i_{Lr}(t_1) = \frac{\begin{pmatrix} -0.5d_p d'_p V_{C_{TP}} + \rho d_p V_{C_{TP}} \\ +0.5d_s d'_s V_{C_{TS}} \end{pmatrix}}{f_{sw}L_r} \tag{6}$$

$$i_{Lr}(t_2) = \frac{\begin{pmatrix} 0.5d_p d'_p V_{C_{TP}} + 0.5d_s d'_s V_{C_{TS}} \\ -(d'_p - \rho)d'_s V_{C_{TS}} \end{pmatrix}}{f_{sw}L_r} \tag{7}$$

$$i_{Lr}(t_3) = \frac{\begin{matrix} 0.5d_p d'_p V_{C_{TP}} \\ -(d_p - d'_s + \rho)d'_p V_{C_{TP}} \\ -0.5d_s d'_s V_{C_{TS}} \end{matrix}}{f_{sw}L_r} \tag{8}$$

In (5) to (8), the expressions of the auxiliary capacitors voltage ($V_{C_{TP}}, V_{C_{TS}}$) are given by (9) and (10):

$$V_{C_{TP}} = \frac{V_{in}}{d'_p} \tag{9}$$

$$V_{C_{TS}} = \frac{V_o}{d'_s} \tag{10}$$

These equations are derived from the volt-second balance equations for the input and output inductor voltages. In these equations, $V_{in}$ is the DC input voltage and $V_o$ is the DC output voltage.

## IV. CI Structure and Current Ripple

### A. CI Structure

As can be seen in Fig. 2, each of the input and output inductor currents ($i_{Lin}(t)$ , $i_{Lo}(t)$) has a DC value ($I_{Lin}, I_{Lo}$) plus an AC ripple ($2\Delta I_{Lin}$, $2\Delta I_{Lo}$) on it. These two types of currents create DC and AC fluxes inside their related magnetic core for each of the $L_{in}$ and $L_o$ which causes core losses and raises the core saturation limit. DC and AC flux cancellation approaches can help reduce the core saturation limit, allowing us to use a smaller core and mitigate core losses. Therefore, integration of $L_{in}$ and $L_o$ could help in this regard.

Fig. 3 shows the IZCC topology with coupling of $L_{in}$ and $L_o$ which are the CI. As shown in this figure, both inductors are wound on a U core, and the sign of the created fluxes from the $L_{in}$ and $L_o$ are in opposite directions and therefore can cancel each other. The created fluxes contain both DC and AC fluxes. The DC fluxes can be easily cancelled, and the AC flux depends on the directions of the current ripple each time. If both currents are increasing or decreasing, the fluxes can cancel each other; otherwise, they add.

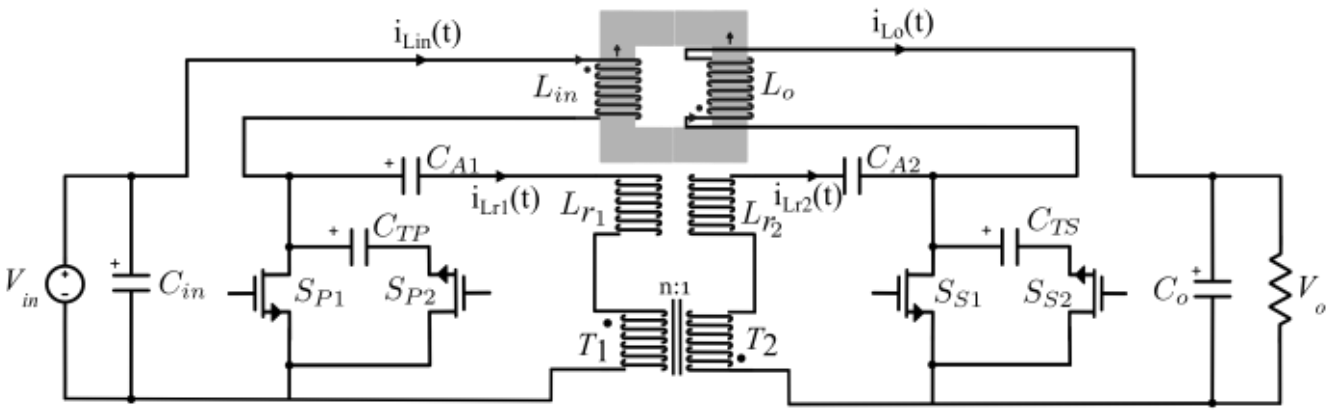

Fig. 3. The CI instructor for $L_{in}$ and $L_o$ in IZCC.

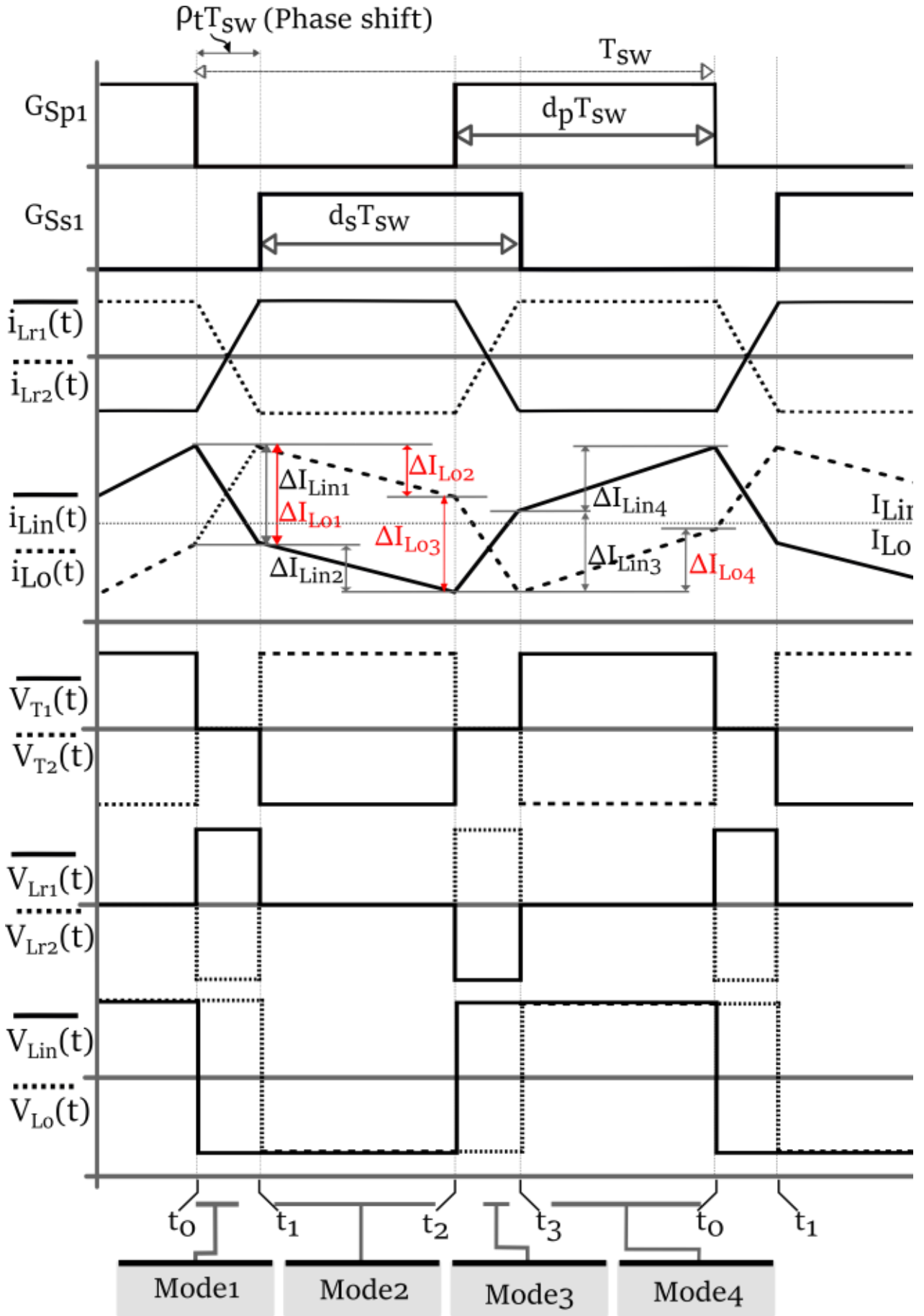

Fig. 4. The IZCC waveforms considering the CI structure

In IZCC, $\rho$ is an essential parameter for the power transfer and output voltage regulation [33]. As shown in Fig. 2, the directions of $i_{Lin}(t)$ and $i_{Lo}(t)$ are not the same in modes 1 and 3, and therefore, their AC fluxes will be additive in these two modes of operation, as shown in the CI in Fig. 3. By considering CI in the IZCC topology, the waveforms of the $i_{Lin}(t)$ and $i_{Lo}(t)$ in Fig. 2 changes to Fig. 4. As can be seen in this figure, the slopes of $i_{Lin}(t)$ and $i_{Lo}(t)$ in modes 1 and 3 are more than the other two modes of operation. This is due to the effect of additive AC fluxes in modes 1 and 3. The additive AC fluxes directly affect current ripples within a short time, thereby increasing the current slope.

### B. CC and Current Ripple in CI

Specifying CC is the key to determining the amount of ripple cancellation or addition in the input and output inductor currents. Fig. 5 shows the equivalent circuit of the CI. By applying Kirchhoff's Voltage Law (KVL) in each input and output side loops, their voltage equations are determined to be

$$\begin{cases} V_{Lin} = L_{in}\dfrac{di_{Lin}}{dt} - L_{io}\dfrac{di_{Lo}}{dt} \\ V_{Lo} = -L_{io}\dfrac{di_{Lin}}{dt} + L_o\dfrac{di_{Lo}}{dt} \end{cases} \tag{9}$$

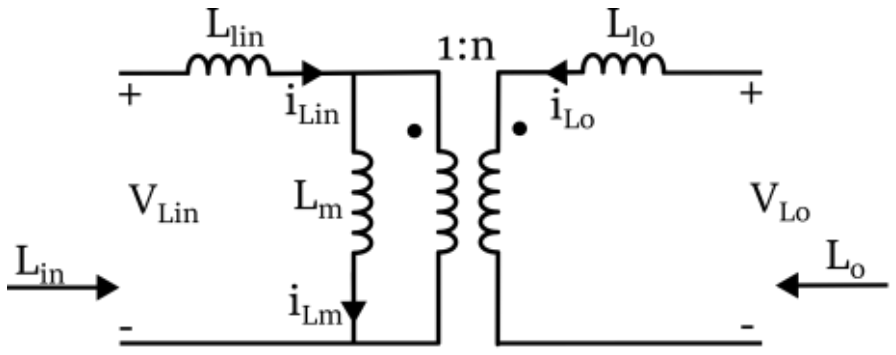

Fig. 5. CI equivalent circuit

By considering a conversion ratio of 1 (i.e., $n = 1$), in Fig. 5, the relation between the mutual inductance ($L_{io}$) and the magnetizing inductance ($L_m$) is as (10). The relations between the input, output open circuit inductances ($L_{in}$, $L_o$), leakage inductances ($L_{lin}$, $L_{lo}$) and $L_m$ are as (11). Equation (12) determines the CC between $L_{in}$ and $L_o$ which is $k_{io}$ [34].

$$L_{io} = nL_m \tag{10}$$

$$\begin{cases} L_{in} = L_m + L_{lin} \\ L_o = L_m + L_{lo} \end{cases} \tag{11}$$

$$k_{io} = \frac{L_m}{\sqrt{L_{in}L_o}} \tag{12}$$

In the symmetrical (balanced) ripple, $L_{in}$=$L_o$=$L$ and, therefore, from (10)-(12), the relations between CI leakage inductances ($L_{lin}$, $L_{lo}$), $k_{io}$, $L_m$, and L are found to be

$$L_m = k_{io}L \tag{13}$$

$$\begin{cases} L_{lin} = (1 - k_{io})L \\ L_{lo} = (1 - k_{io})L \end{cases} \tag{14}$$

Therefore, from (9)-(14), the state space equations for the CI can be driven by (15).

$$\begin{bmatrix} L & -k_{io}L \\ -k_{io}L & L \end{bmatrix} \begin{bmatrix} \dfrac{di_{Lin}}{dt} \\ \dfrac{di_{Lo}}{dt} \end{bmatrix} = \begin{bmatrix} V_{Lin} \\ V_{Lo} \end{bmatrix} \tag{15}$$

From (15), the input and output inductor currents' slopes can be determined to be

$$\begin{cases} \dfrac{di_{Lin}}{dt} = \dfrac{1}{L(1 - k_{io}^2)}(V_{Lin} + k_{io}V_{Lo}) \\ \dfrac{di_{Lo}}{dt} = \dfrac{1}{L(1 - k_{io}^2)}(V_{Lo} + k_{io}V_{Lin}) \end{cases} \tag{16}$$

As can be seen in Fig. 4, $i_{Lin}$ and $i_{Lo}$ waveforms are linear and, therefore, from (16), the input and output inductor current ripples ($\Delta i_{Lin}, \Delta i_{Lo}$) in each mode of the operation can be found as (17). In these equations, $\Delta t$ is the time duration of each operating mode.

$$\begin{cases} \Delta i_{Lin} = \dfrac{\Delta t}{L(1-k_{io}^2)}(V_{Lin}+k_{io}V_{Lo}) \\ \Delta i_{Lo} = \dfrac{\Delta t}{L(1-k_{io}^2)}(V_{Lo}+k_{io}V_{Lin}) \end{cases} \tag{17}$$

As shown in Fig. 5, the relation between $i_{Lin}$, $i_{Lo}$ and the magnetizing current ($i_{Lm}$) is as (18). In this equation, the conversion ratio is one ($n = 1$).

$$i_{Lm} = i_{Lin} + i_{Lo} \tag{18}$$

From (18) and (17) the magnetizing current ripple ($\Delta i_{Lm}$) is obtained as (19).

$$\Delta i_{Lm} = \frac{\Delta t}{L(1-k_{io})}(V_{Lin}+V_{Lo}) \tag{19}$$

As seen from (17) and (19), the ripples are related to the $k_{io}$, and by properly adjusting the $k_{io}$, the desired current ripples can be achieved. As can be seen in Fig. 4, the input and output inductors' currents have different ripples in each mode of operation. It can be confirmed from (17) as well. From (9), (10) and Fig. 4, the values of the $V_{Lin}$ and $V_{Lo}$ in each mode of the operation are as (20) and (21). And the value of $\Delta t$ in each mode from Fig. 4 is as (22).

$$\begin{cases} V_{Lin} = -\dfrac{d_p}{d'_p}V_{in} & [Modes\ 1,2] \\ V_{Lin} = V_{in} & [Modes\ 3,4] \end{cases} \tag{20}$$

$$\begin{cases} V_{Lo} = -\dfrac{d_s}{d'_s}V_o & [Modes\ 1,4] \\ V_{Lo} = V_o & [Modes\ 2,3] \end{cases} \tag{21}$$

$$\Delta t = \begin{cases} \rho T_{sw} & [Modes\ 1] \\ (d'_p - \rho)T_{sw} & [Modes\ 2] \\ (d_p - d'_s + \rho)T_{sw} & [Modes\ 3] \\ (d'_s - \rho)T_{sw} & [Modes\ 4] \end{cases} \tag{22}$$

Fig. 6 shows the effect of $k_{io}$ on the input, output, and magnetizing inductor current ripples in four modes of operation for the IZCC. As can be seen in this figure, by increasing the value of $k_{io}$, the ripple currents for modes 2 and 4 decrease in both input and output inductor currents, whereas they increase for modes 1 and 3 in both currents. Thus, using only the CI in IZCC cancels the AC flux in two modes of operation and adds it in the other two; therefore, core losses cannot be mitigated.

## V. IM Structure and Design

### A. Integration of CI and Resonant Inductors

As mentioned in the section IV part A, the AC fluxes are additive in modes 1 and 3. Fig. 4 shows that the resonant inductor's voltage ($V_{Lr1}(t)$, $V_{Lr2}(t)$) have their peak values during these two modes, and also, the resonant inductor's current ($i_{Lr1}(t)$, $i_{Lr2}(t)$) varies from positive to negative or vice versa. Therefore, these two inductors can be used to cancel flux and thus ripple during modes 1 and 3. Fig. 7 shows the IZCC topology with the CI and RIs integrated on a core. Thus, the ripple cancellation can be achieved in all modes of operation.

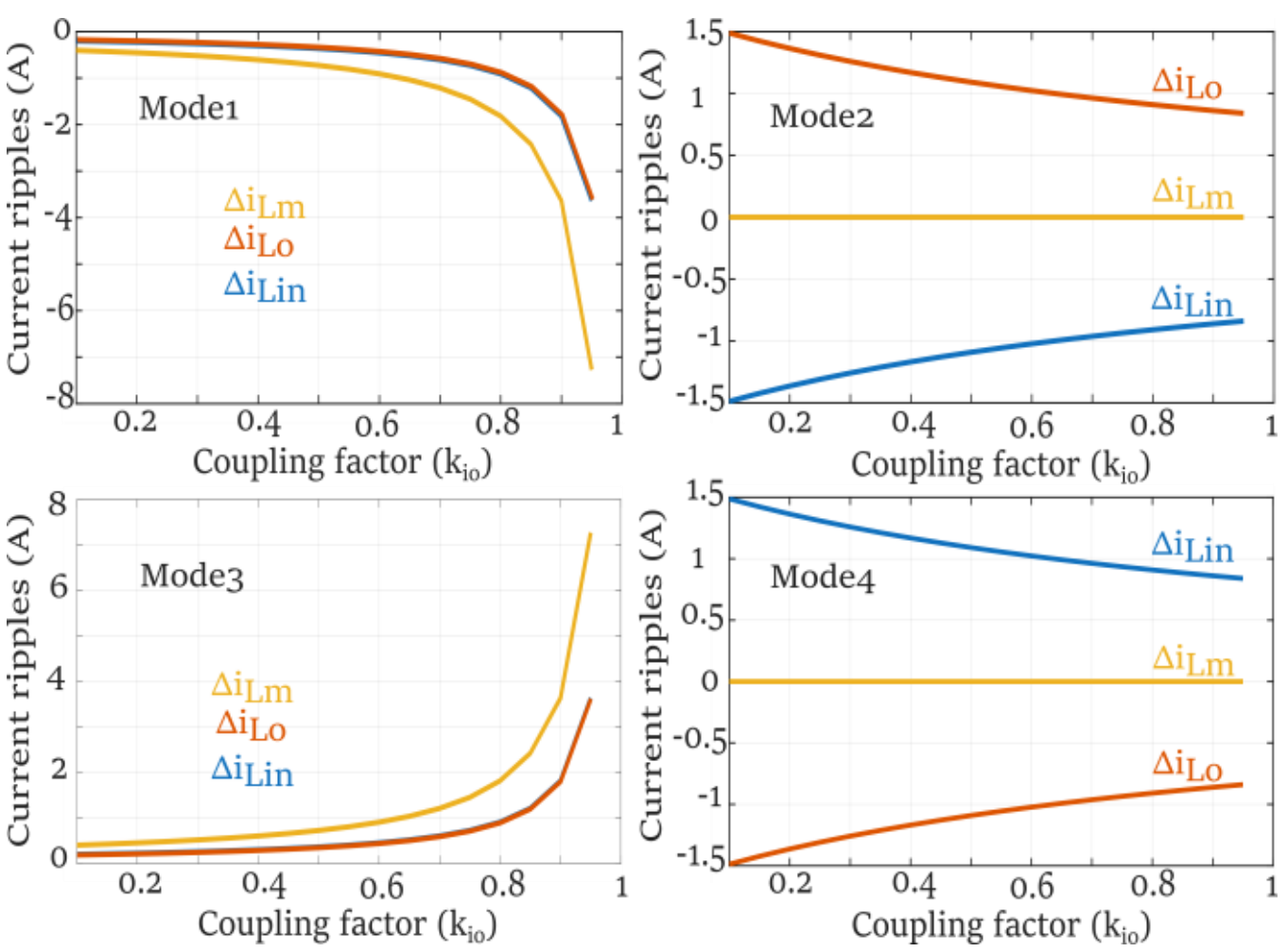

Fig. 6. Input and output inductor current ripple variations based on the coupling coefficient (k) changes

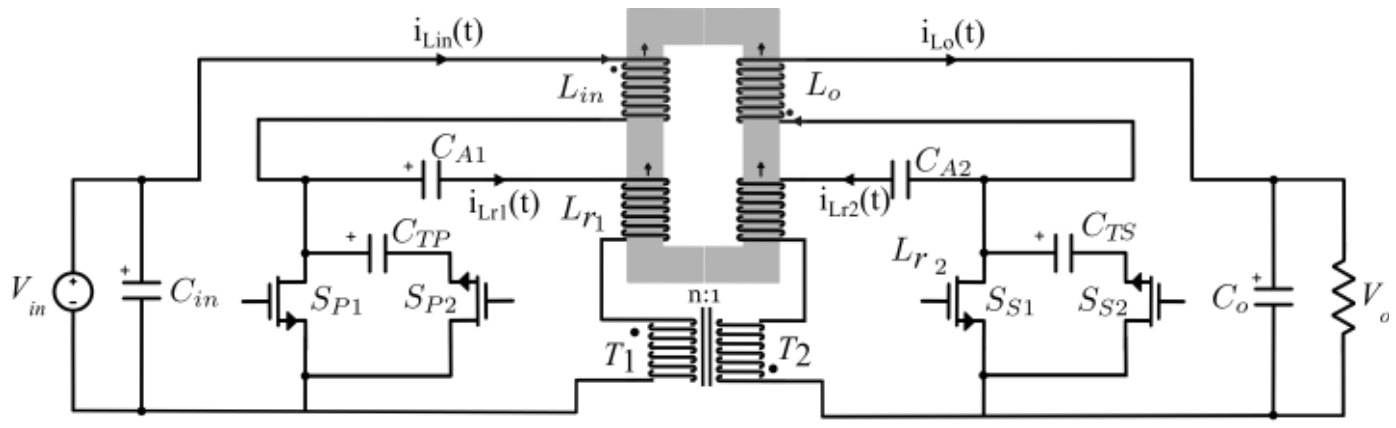

Fig. 7. IZCC topology with the integration of CI and Ris

### B. Full IM Structure

Although the RIs help in ripple cancellation at modes 1 and 3, they are additive at modes 2 and 4, and their flux can affect the CI's operation at those modes. Therefore, additional windings should be used to mitigate the effect of RIs in modes 2 and 4. In the Full IM structure, the transformer is also integrated with CI and RIs in this regard. Fig. 8 shows the IZCC topology with the IM structure. In the IZCC topology, the transformer must be designed with a high-value CC between the primary and secondary windings, and the RIs are then designed separately to provide flexibility for winding relocation and arrangement.

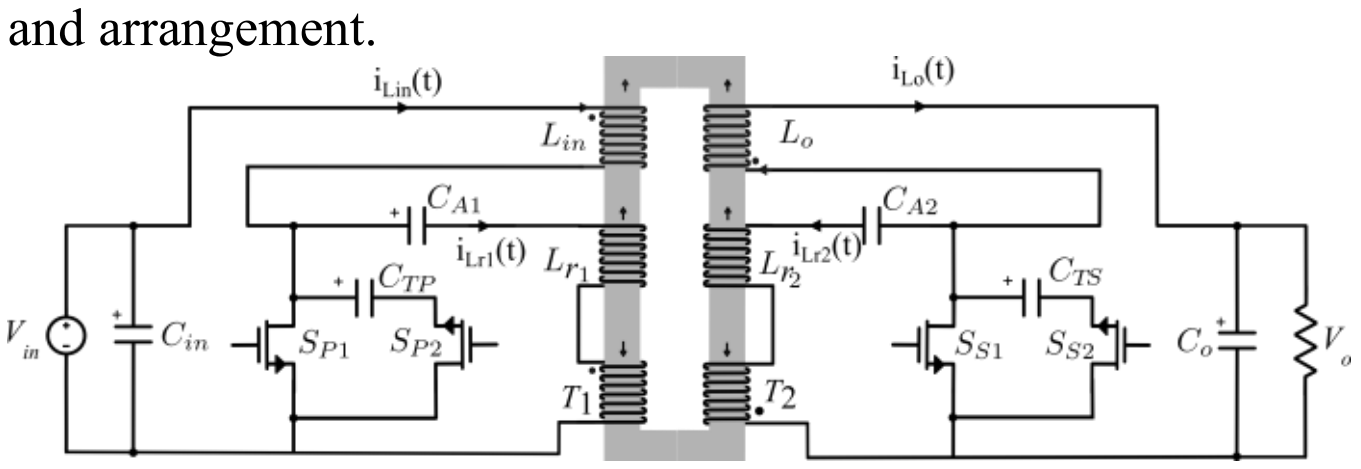

Fig. 8. IZCC topology with IM structure

Based on the IZCC operation, the sign of the CCs between every two windings should be as Table I. The IM winding arrangements should be such that the sign of the CCs between every two windings is as shown in this table. These signs come from the values of the current and the voltages from Fig. 4 and the necessity of some signs to be positive or negative. For example, the CC sign for the CI or the transformer windings should be negative to have flux cancellation. The value of each CC depends on the flux and ripple cancellations required

TABLE I. THE CC SIGNS BETWEEN EVERY TWO WINDINGS OF THE IM.

| windings | CC sign | windings | CC sign | windings | CC sign |
|---|---|---|---|---|---|
| $L_{in}, L_o$ | $K_{io} < 0$ | $L_o, L_{r2}$ | $K_{or2} > 0$ | $L_{r1}, L_{T1}$ | $K_{r1T1} < 0$ |
| $L_{in}, L_{r1}$ | $K_{ir1} > 0$ | $L_o, L_{T2}$ | $K_{oT2} < 0$ | $L_{r1}, L_{T2}$ | $K_{r1T2} > 0$ |
| $L_{in}, L_{T1}$ | $K_{iT1} < 0$ | $L_o, L_{r1}$ | $K_{or1} < 0$ | $L_{r2}, L_{T1}$ | $K_{r2T1} > 0$ |
| $L_{in}, L_{r2}$ | $K_{ir2} < 0$ | $L_o, L_{T1}$ | $K_{oT1} > 0$ | $L_{r2}, L_{T2}$ | $K_{r2T2} < 0$ |
| $L_{in}, L_{T2}$ | $K_{iT2} > 0$ | $L_{r1}, L_{r2}$ | $K_{r1r2} < 0$ | $L_{T1}, L_{T2}$ | $K_{T1T2} < 0$ |

### C. IM Design

The IZCC converter in this paper is intended to operate at a high frequency. Therefore, achieving soft switching for the switches is critical. As shown in the ZVS conditions in (1) to (4), achieving ZVS for the switches relies on the input and output inductor current ripples and the resonant inductor current. Thus, a minimum value of ripples is required for $i_{Lin}(t)$ and $i_{Lo}(t)$. Therefore, the IM design should be such that the current ripple cancellation is limited under ZVS conditions [31][33]. Table II shows the IZCC design parameters. Based on these parameters, the initial values of the resonant inductor current can be obtained using (5)-(8). Having the values of $I_{Lin}$ and $I_{Lo}$, the switch output capacitor, from (9) and (10), the minimum $\Delta I_{Lin}$ and $\Delta I_{Lo}$ can be found in (1) to (4). Therefore, the minimum value of $\Delta I_{Lin}$ and $\Delta I_{Lo}$ are as $\Delta I_{Lin}$ = $\Delta I_{Lo}$=2.5A.

TABLE II. IZCC DESIGN PARAMETERS

| Parameter | Value | Parameter | Value |
|---|---|---|---|
| $V_{in}$ | 170 $V_{dc}$ | $C_{TP} = C_{TS}$ | 3 $\mu$F |
| $V_o$ | 170 $V_{dc}$ | $d_P$ | 0.5 |
| Output power ($P_o$) | 0.25-0.5 kW | $d_S$ | 0.5 |
| $f_{sw}$ | 150 kHz | $\rho_t$ | 0.01 – 0.2 |
| $C_{A1} = C_{A2}$ | 3 $\mu$F | n | 1 |

In the IM operation, all windings are coupled together, and therefore, the windings have an effect on each other. Therefore, specifying the proper value of the CCs is crucial in this regard. Equation (23) is the state space equation and (24) is the expanded form of the state space equation.

$$V_W = L_W.\frac{d}{dt}I_W \tag{23}$$

In (24), the parameters $N_{in}$, $N_o$, $N_{r1}$, $N_{r2}$, $N_{T1}$, $N_{T2}$ are the number of turns for input, output, primary resonant inductor, secondary resonant inductor, primary transformer, and secondary transformer windings, respectively. This equation is considered for a U magnetic core; therefore, the reluctance $R$ is the same for all the windings. By specifying the inductance of each winding, the CC values can be determined.

In this paper, the RIs are considered separated from the transformer, and therefore, the transformer needs to be designed with a high CC value with a high magnetizing inductance value ($L_m$). The values of the $L_{in}$ and $L_o$ are specified in such a way that the minimum considered power transfer could transfer such that the CI current ripples do not exceed the minimum average current. It prevents the input and output inductor currents from becoming negative, thereby avoiding circulating current. The value of the RIs is considered based on the ZVS conditions in (1) to (4) and the power transfer [33]. Therefore, the primary values of the inductances are as Table III.

TABLE III. THE PRIMARY CALCULATED VALUE OF THE MAGNETICS IN IZCC

| Inductor | Value | Inductor | Value |
|---|---|---|---|
| $L_{in}$ | 230 $\mu$H | $L_{lr2}$ | 65 $\mu$F |
| $L_o$ | 230 $\mu$H | $L_{T1}$ | 1000 $\mu$F |
| $L_{lr1}$ | 65 $\mu$H | $L_{T2}$ | 1000 $\mu$F |

In IZCC, the proper value of the RIs is critical in order to transfer the desired power to the load and achieve soft switching [31]. By integrating the RIs with the other magnetics, the other windings are also coupled with the RIs, and the equivalent circuit for the integrated RIs will be as Fig. 9. Fig. 9(a) shows the coupled RIs and Fig. 9(b) shows the T model of the integrated RIs. The primary and secondary leakage inductances ($L_{lr1}$, $L_{lr2}$) are the equivalent leakage inductance coupled with all other windings. Therefore, they can be found as (25) and (26). In these equations, the sign of each inductance comes from the sign of the CCs specified in Table I. Parameter $L_{r12}$ is the mutual inductance between the resonant inductors' windings, and $L_{ir1}$, $L_{or1}$, $L_{r1T1}$, and $L_{r1T2}$ are the mutual inductances between the $L_{r1}$ and the input, output, primary transformer, and secondary transformer windings, respectively. Also, and $L_{ir2}$, $L_{or2}$, $L_{r2T1}$, $L_{r2T2}$ are the mutual inductances between $L_{r2}$ and the input, output, primary transformer, and secondary transformer windings, respectively. As the IZCC in this paper operates one by one and the structure is symmetric, the effect of CI and the transformer windings on the value of $L_{lr1}$, $L_{lr2}$ are not significant because they cancel each other out in (25) and (26). Thus, the value of self-inductances and $L_{r12}$ are critical to specify $L_{lr1}$ and $L_{lr2}$.

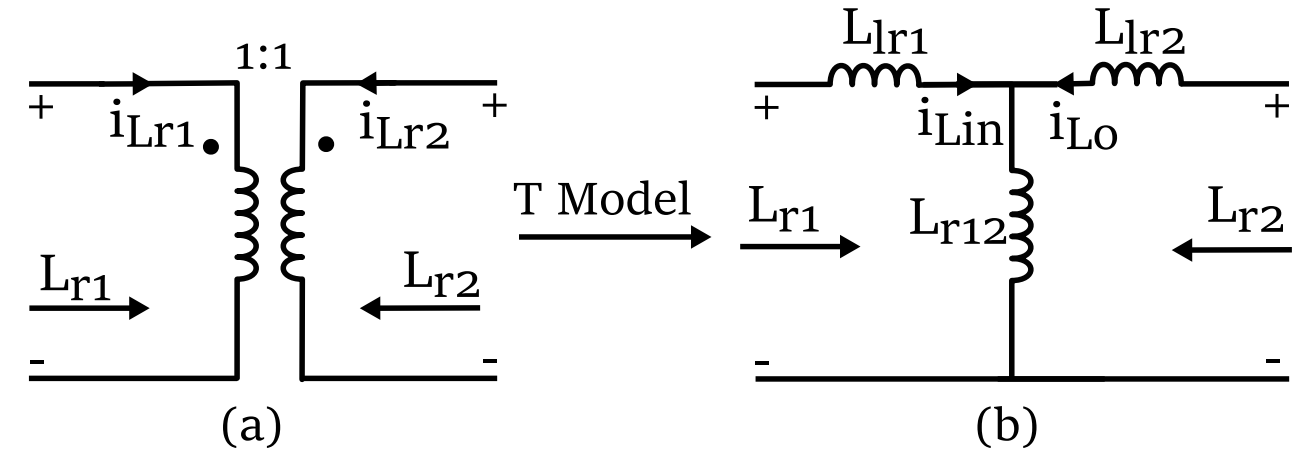

Fig. 9. The integrated RIs equivalent circuit. (a) The coupled RIs circuit. (b) Equivalent T model circuit.

$$\begin{bmatrix} V_{Lin}(t) \\ V_{Lo}(t) \\ V_{Lr1}(t) \\ V_{Lr2}(t) \\ V_{T1}(t) \end{bmatrix} = \frac{1}{R} \begin{bmatrix} N_{in}^2 & k_{io}N_{in}N_o & k_{ir1}N_{in}N_{r1} & k_{ir2}N_{in}N_{r2} & k_{iT1}N_{in}N_{T1} & k_{iT2}N_{in}N_{T2} \\ k_{io}N_{in}N_o & N_o^2 & k_{or1}N_oN_{r1} & k_{or2}N_oN_{r2} & k_{oT1}N_oN_{T1} & k_{oT2}N_oN_{T2} \\ k_{ir1}N_{in}N_{r1} & k_{or1}N_oN_{r1} & N_{r1}^2 & k_{r1r2}N_{r1}N_{r2} & k_{r1T1}N_{r1}N_{T1} & k_{r1T2}N_{r1}N_{T2} \\ k_{ir2}N_{in}N_{r2} & k_{or2}N_oN_{r2} & k_{r1r2}N_{r1}N_{r2} & N_{r2}^2 & k_{r2T1}N_{r2}N_{T1} & k_{r2T2}N_{r2}N_{T2} \\ k_{iT1}N_{in}N_{T1} & k_{oT1}N_oN_{T1} & k_{r1T1}N_{r1}N_{T1} & k_{r2T1}N_{r2}N_{T1} & N_{T1}^2 & k_{T1T2}N_{T1}N_{T2} \end{bmatrix} \frac{d}{dt} \begin{bmatrix} i_{Lin}(t) \\ i_{Lo}(t) \\ i_{Lr1}(t) \\ i_{Lr2}(t) \\ i_{T1}(t) \end{bmatrix} \tag{24}$$

$$L_{lr1} = L_{r1} - L_{r12} + L_{ir1} - L_{or1} - L_{r1T1} + L_{r1T2} \quad (25)$$
$$L_{lr2} = L_{r2} - L_{r12} - L_{ir2} + L_{or2} + L_{r2T1} - L_{r2T2} \quad (26)$$

For transferring the desired power to the load, the values of resonant self-inductances and the RIs coupling coefficient should be specified such that the leakage values in Table III are found.

### D. Core Selection and Winding Arrangements

Given the inductance values and core geometry, the number of turns and required air gap could be determined. The core geometry should be selected so that by properly arranging the windings, both DC and AC flux cancellations are feasible. In this case, the U core is the one for which DC flux cancellation for the CI can be easily achieved by negatively coupling the CI windings, and the desired AC flux cancellation can be achieved by properly relocating and arranging the windings. In this regard, the UR Ferrite-7070-UR81 (81-42-16) core is selected. With the core specification, the number of turns could be easily determined. Table IV shows the number of turns for the CI, transformer, airgap, and core specifications of the selected core. The number of turns for the RIs comes from the winding arrangement and the value of the $k_{r1r2}$.

TABLE IV. DESIGN PARAMETERS BASED ON UR FERRITE-7070- UR81 CORE.

| Parameter | Value | Parameter | Value |
|---|---|---|---|
| core volume | 50.023 mm$^3$ | airgap | 0.3 mm |
| $N_{in}$ | 21 | $N_{T1}$ | 45 |
| $N_o$ | 21 | $N_{T2}$ | 45 |

Fig. 10(a) shows the IM's winding arrangement on the U core. In this, the transformer windings are arranged such that they are perfectly coupled. The CI windings are arranged so that their CC can be adjusted to provide sufficient ripple to achieve ZVS. And the RIs' windings are located such that their CC is as small as possible to generate the leakage value as much as the values in Table III. The direction of the fluxes in this figure is specified based on the CC signs in Table I.

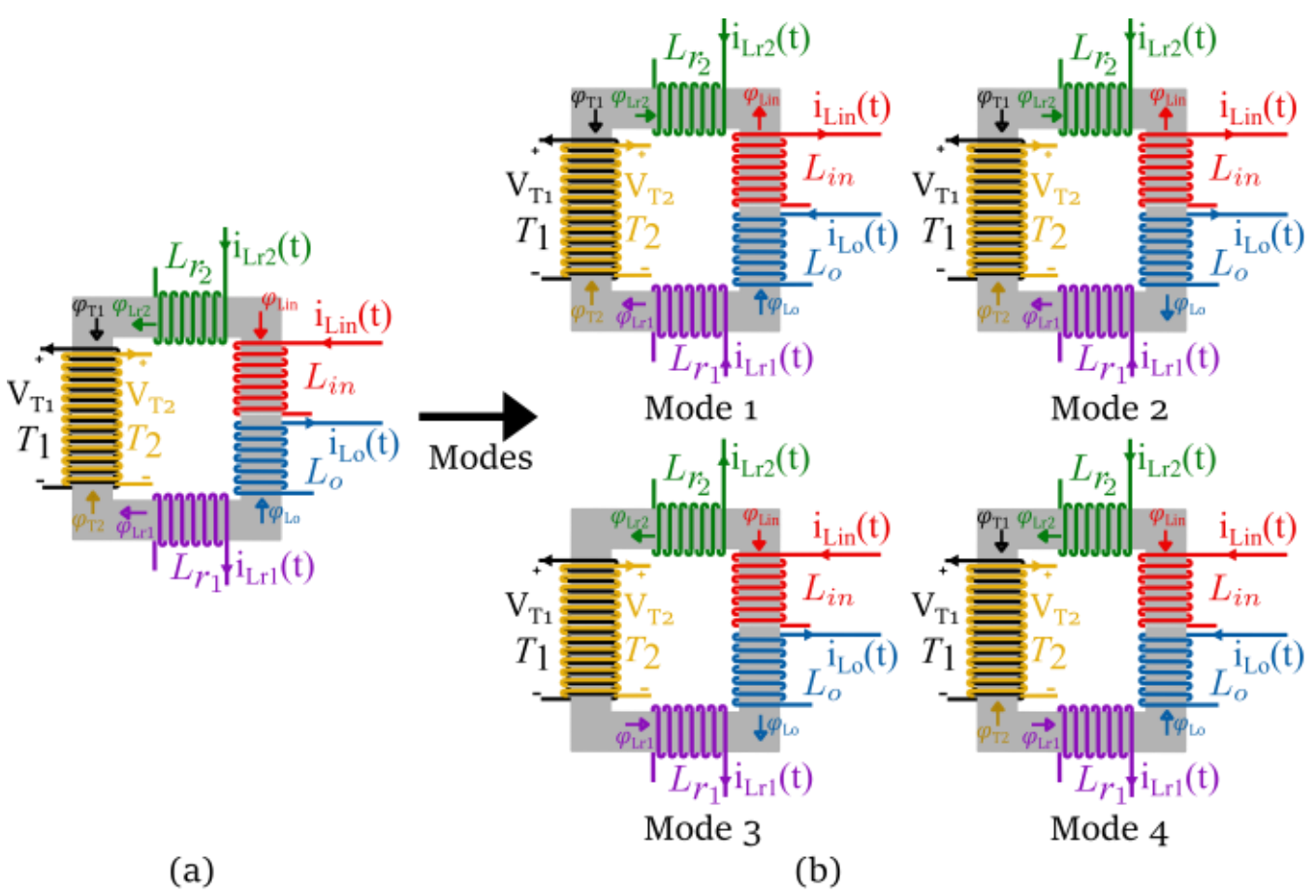

Fig. 10. The winding arrangement and flux directions in each mode for the full IM on the U core. (a) Winding arrangement. (b) AC flux direction in each mode of operation.

Fig. 10(b) shows the flux directions in each mode of operation. As can be seen in this figure, the flux directions for the CI are in the same direction in modes 1 and 3, where in these two modes, the RIs fluxes are opposite to the CI fluxes, and the RIs could cancel the CI flux in these two modes of operation. As the transformer flux is the integral of the winding voltages, the direction of the transformer flux does not change during the modes. Except in mode 3, where the voltage value of the windings is zero in Fig. 4 and therefore the windings' fluxes go to zero. In mode 1, the transformer fluxes retain their values from the previous mode of operation. As the transformer windings are perfectly coupled, the primary and secondary windings are cancelling each other's flux, and only a small magnetizing flux will remain, which does not significantly impact the operation of the other windings.
The IM structure in

Fig. 10(a) is simulated in ANSYS based on the values in Table III and Table IV. The number of turns for the primary and secondary resonant self-inductors ($L_{r1}$, $L_{r2}$) are adjusted such that the values of the $L_{lr1}$, $L_{lr2}$ become as the values of the Table III. Therefore, the RIs parameter values are as Table V.

TABLE V. THE OBTAINED PARAMETER VALUES FOR THE RIS AFTER SIMULATION

| Parameter | Value | Parameter | Value |
|---|---|---|---|
| $N_{r1}$ | 22 | $L_{r1}$ | 257 µH |
| $N_{r2}$ | 22 | $L_{r2}$ | 257 µH |

After simulation of the IM, the core flux density (B) distribution in each mode of operation is as

Fig. 11. As shown in this figure, the highest flux density is concentrated around the RIs windings, and proper flux cancellation occurs in the CI and the transformer.

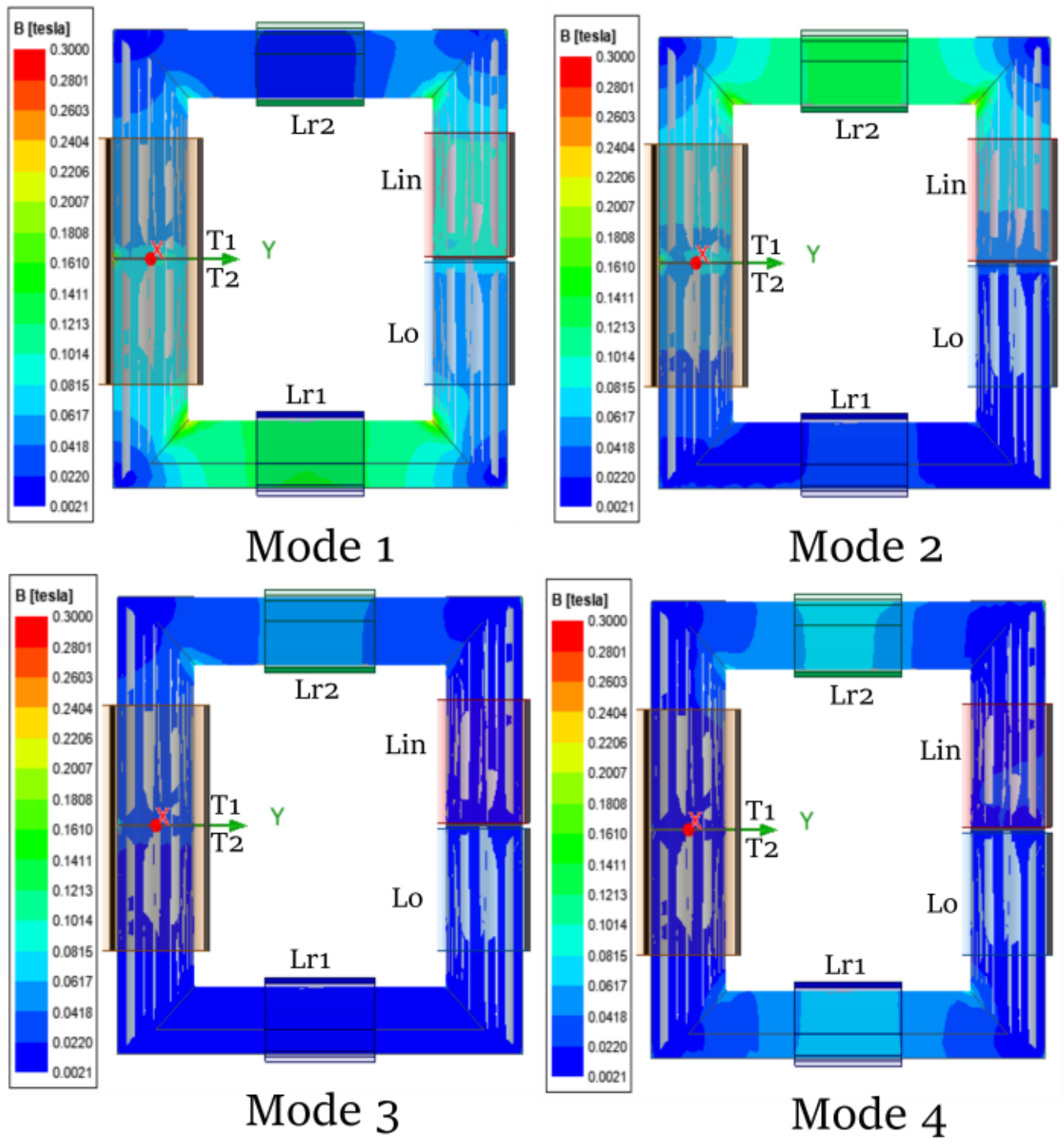

Fig. 11. The flux density distribution for the designed IM in different modes of operation.

## VI. HARDWARE RESULTS

A 0.5 kW IM hardware has been developed based on the specifications in Table II, Table III, Table IV, and Table V. The hardware setup with the IM is shown in Fig. 12. The winding arrangements in IM are based on the arrangement shown in Fig. 10. The IZCC setup consists of three types of printed circuit boards (PCBs). The first board is the daughter card, which mounts the switches and gate drivers. The utilized switches are gallium nitride (GaN) from Infineon and are identified by the part number GS66516T. GaN switches are attractive for high-frequency applications due to their fast-switching speeds, compact size, and lack of reverse recovery losses. The third board is a Texas Instruments (TI) microcontroller-unit (MCU) board with part number TMS320F28379D, which generates the PWM for the GaN switches.

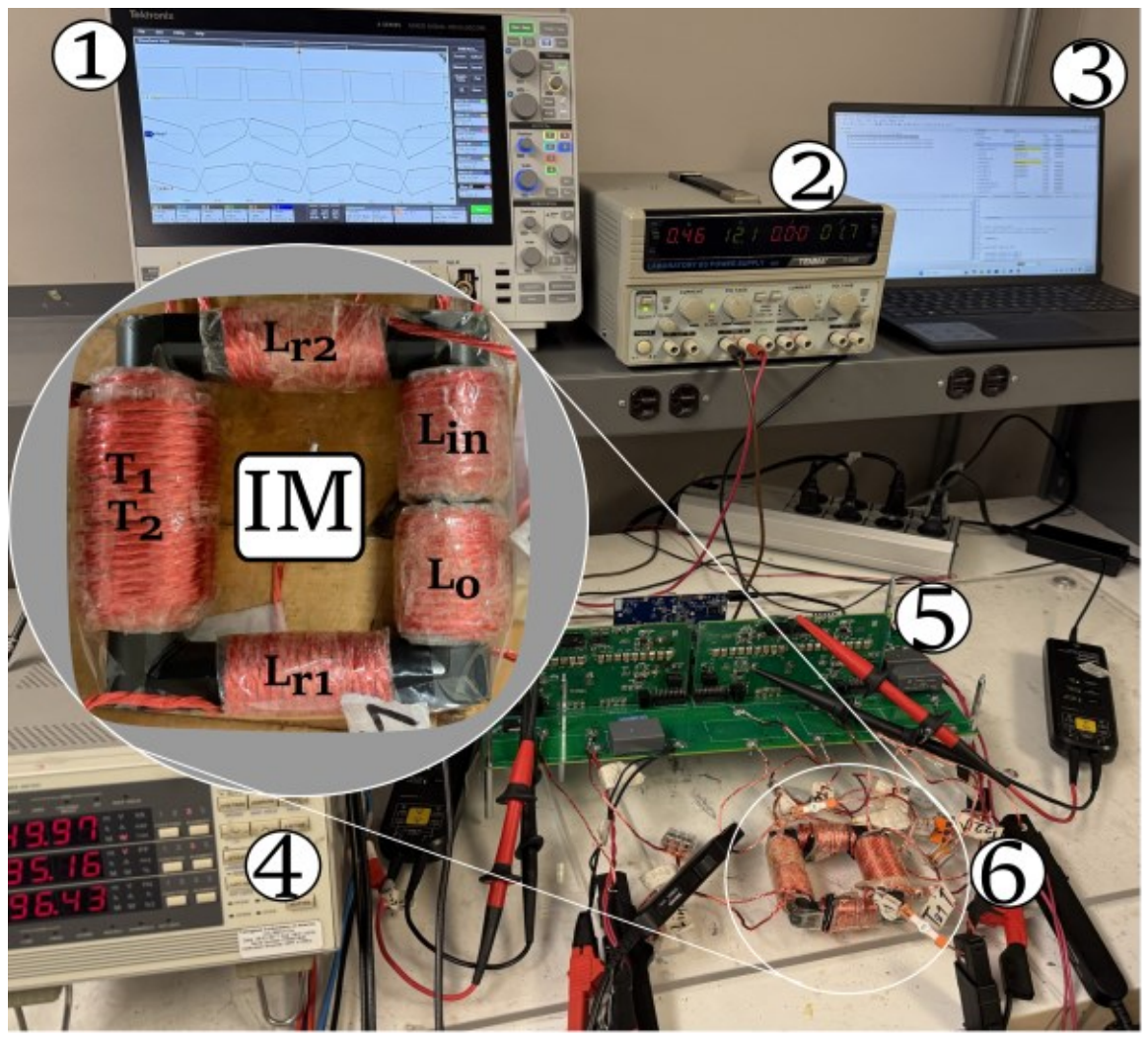

Fig. 12. Hardware setup for the proposed IM with the IZCC converter containing 1) oscilloscope, 2) auxiliary DC power supply, 3) computer for programming, 4) power analyzer, 5) IZCC hardware board, 6) the designed IM.

The measured CC values between every two windings in the implemented IM are as Table VI. The sign of each CC is based on the signs in Table I. As can be seen in this table, the CC value between $L_{in}$ and $L_o$ which is $k_{io}$= 0.8733. This value is specified to ensure ZVS across all switches. Also, the CC values between the CI and RIs are not close to 1 because of the need to maintain a minimum ripple in the phase-shift modes for ZVS. The CC value for the transformer is close to 1 to achieve maximum flux cancellation and minimize the magnetizing flux. The number of turns for each winding is as Table IV and Table V. The measured values in the built IM are as Table VII. From these measured inductances and the CC values, the measured value of the resonant leakage inductances can be obtained as $L_{lr1}$ = 64.5 $\mu$H, $L_{lr2}$ = 64.85 $\mu$H. These values are very close to the simulated values in Table III.

TABLE VI. MEASURED COUPLING COEFFICIENT BETWEEN EVERY TWO WINDINGS AT THE DESIGNED IM.

| CC | Value | CC | Value | CC | Value |
|---|---|---|---|---|---|
| $k_{io}$ | 0.8733 | $k_{or2}$ | 0.7713 | $k_{r1T1}$ | 0.8271 |
| $k_{ir1}$ | 0.7732 | $k_{oT2}$ | 0.7597 | $k_{r1T2}$ | 0.8271 |
| $k_{iT1}$ | 0.7547 | $k_{or1}$ | 0.8389 | $k_{r2T1}$ | 0.8195 |
| $k_{ir2}$ | 0.8322 | $k_{oT1}$ | 0.7589 | $k_{r2T2}$ | 0.8195 |
| $k_{iT2}$ | 0.7553 | $k_{r1r2}$ | 0.7492 | $k_{T1T2}$ | 0.9984 |

TABLE VII. MEASURED INDUCTANCE VALUES FOR EACH WINDING OF THE IMPLEMENTED IM.

| Inductance | Value ($\mu$H) | Inductance | Value ($\mu$H) |
|---|---|---|---|
| $L_{in}$ | 230 | $L_{r2}$ | 258.88 |
| $L_o$ | 232.3 | $L_{T1}$ | 901.5 |
| $L_{r1}$ | 257.1 | $L_{T2}$ | 903.1 |

Fig. 13 shows the current waveforms of the CI and RIs, along with the gate-to-source voltages of the primary and secondary main switches. This figure shows that the $i_{Lin}(t)$ and $i_{Lo}(t)$ have ripples that, as mentioned, are necessary in order to achieve soft switching on all switches in IZCC, and these waveforms are in the same shape as the waveforms shown in Fig. 4.

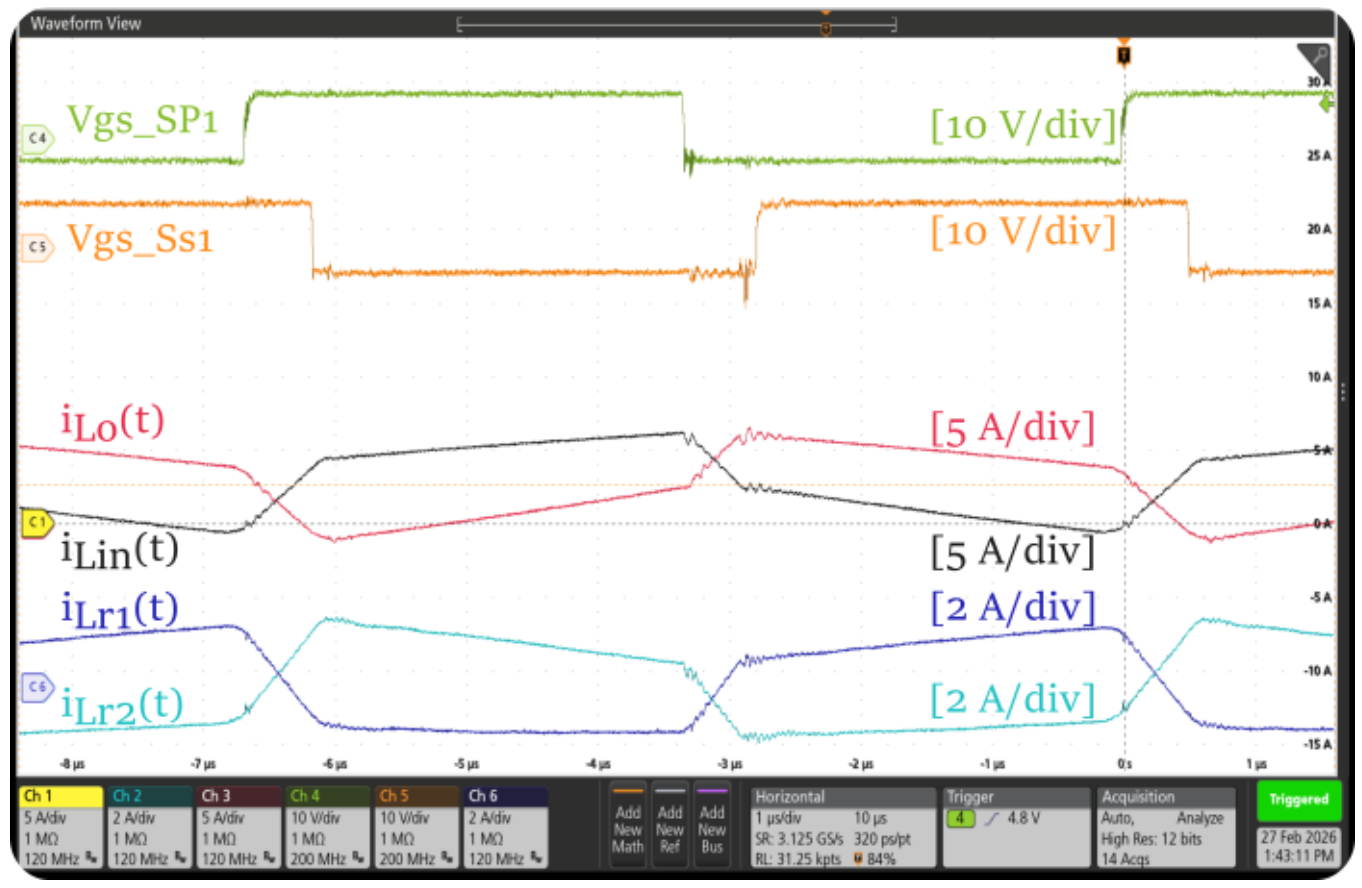

Fig. 13. IM waveforms with the switches' gate-to-source voltages.

Fig. 14 shows the CI and RIs waveforms with the transformer voltages. As the transformer windings are in series connection with the RIs windings, their current waveforms are the same as the RIs current waveform.

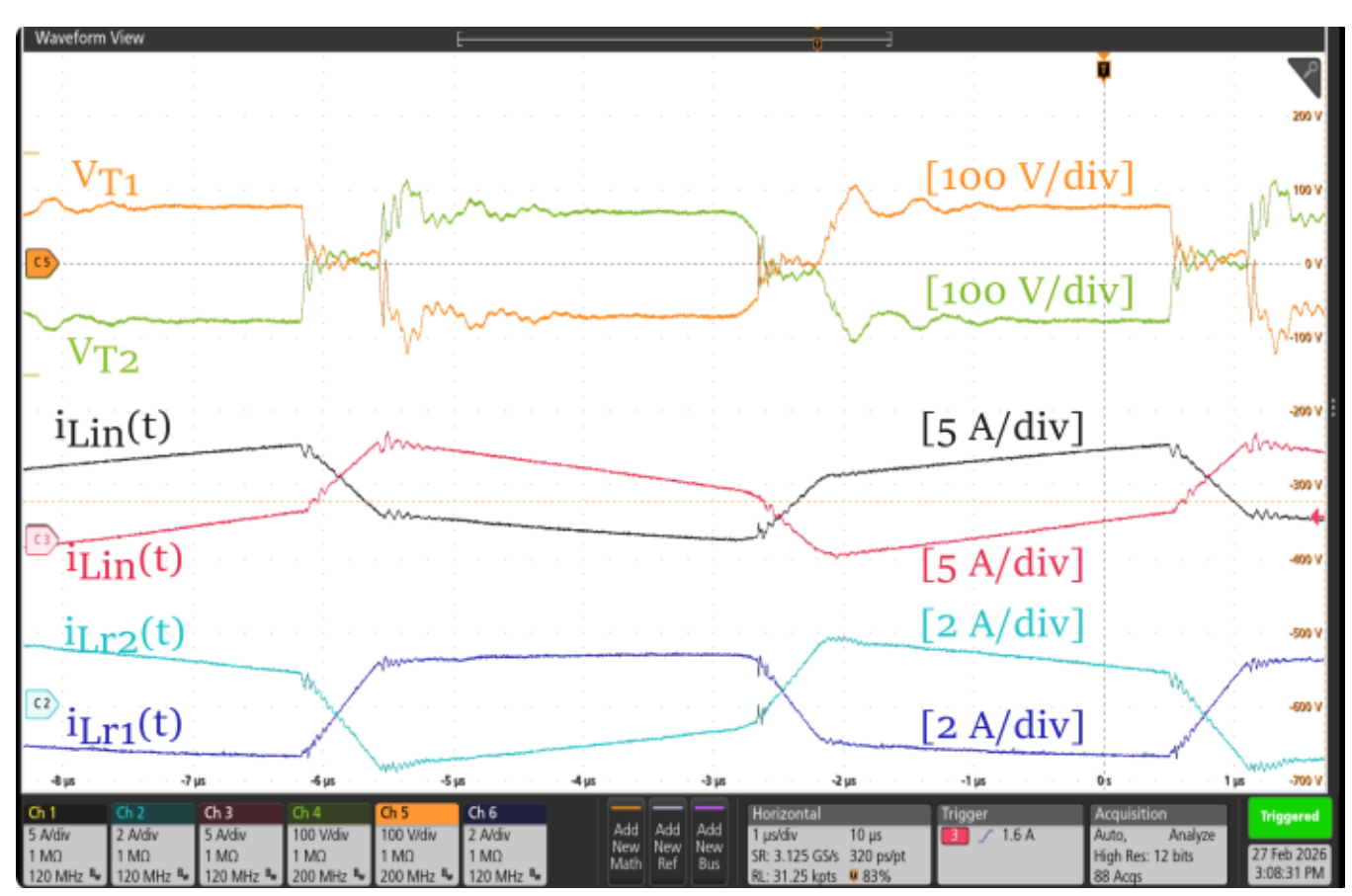

Fig. 14. IM waveforms with the transformer voltage waveforms

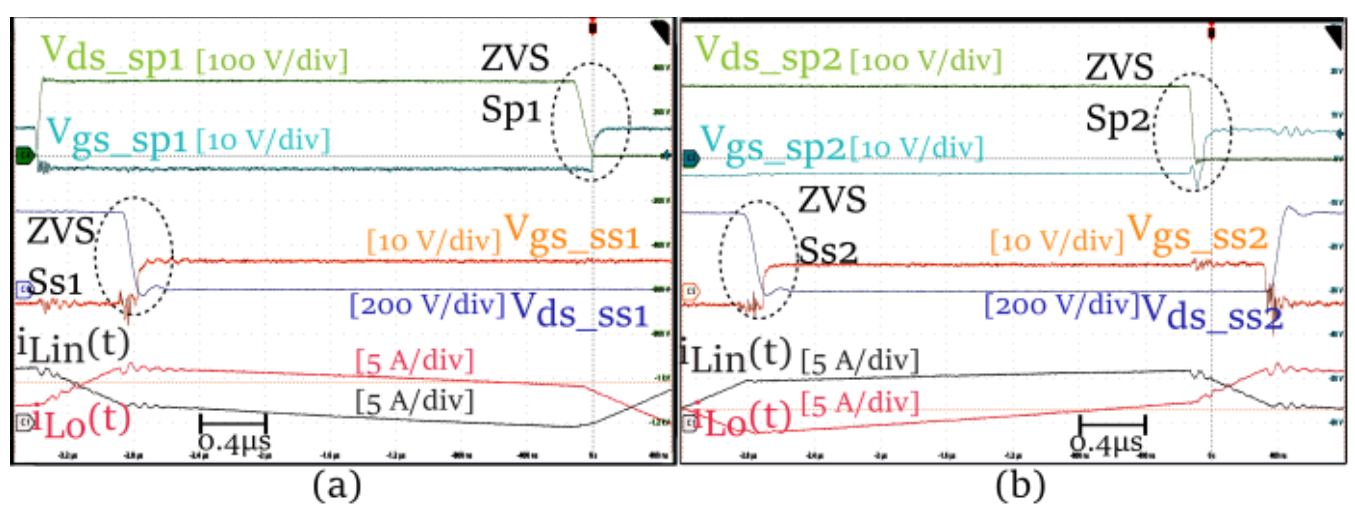

Fig. 15. ZVS realization for the primary- and secondary-side switches using IM in IZCC. (a) ZVS status for $S_{p1}$ and $S_{s1}$ switches (b) ZVS status for $S_{p2}$ and $S_{s2}$ switches.

Fig. 16Fig. 15 shows the ZVS realization for the switches on the primary and secondary sides. Fig. 15(a) shows the ZVS status of the primary and secondary main switches ($S_{p1}$ and $S_{s1}$). And Fig. 15(b) shows the ZVS status of the primary and secondary auxiliary switches ($S_{p2}$ and $S_{s2}$). As shown in these figures, all switches can achieve ZVS using the implemented IM.

Fig. 16 shows the transient response of the IZCC using the IM. Fig. 16(a) shows the transient response when the load suddenly changes from half to full load. Fig. 16(b) shows the transient response from full load to half load. As shown, the converter can quickly regulate the output voltage when a sudden change in the load occurs.

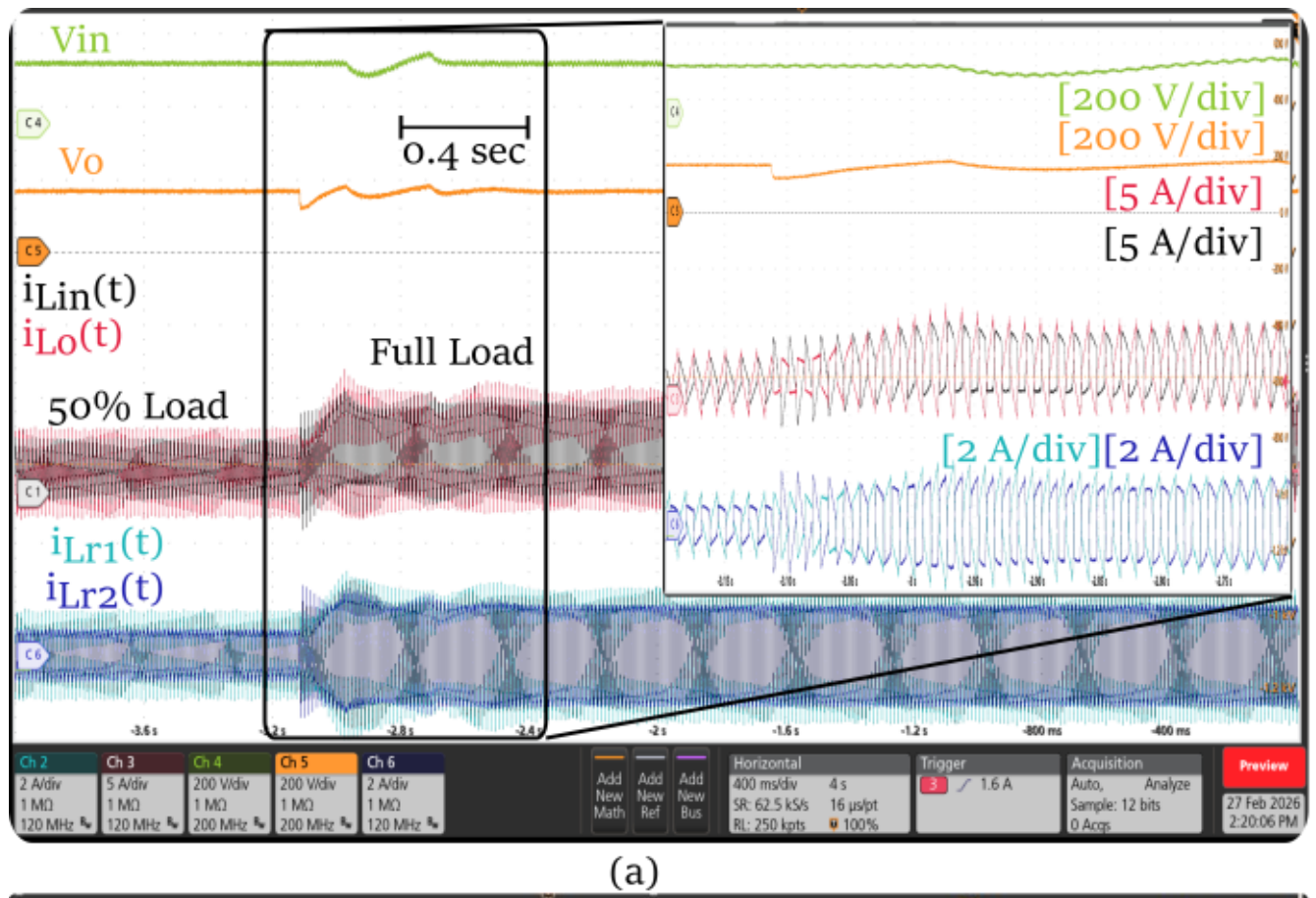

(a)

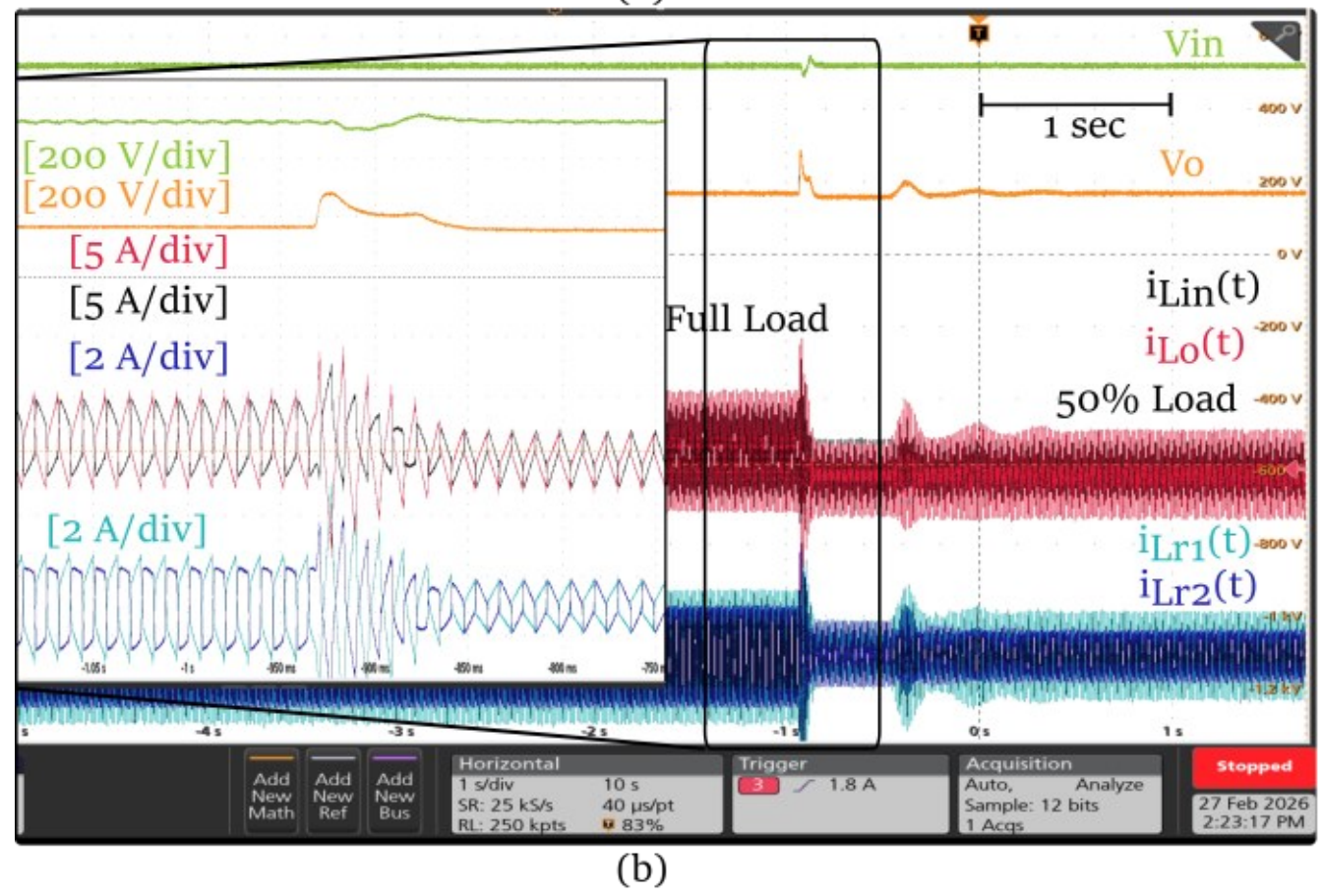

(b)

Fig. 16. The transient response considering the designed IM in the IZCC. (a) Changing from 50% load to full load. (b) Changing from full load to 50% load.

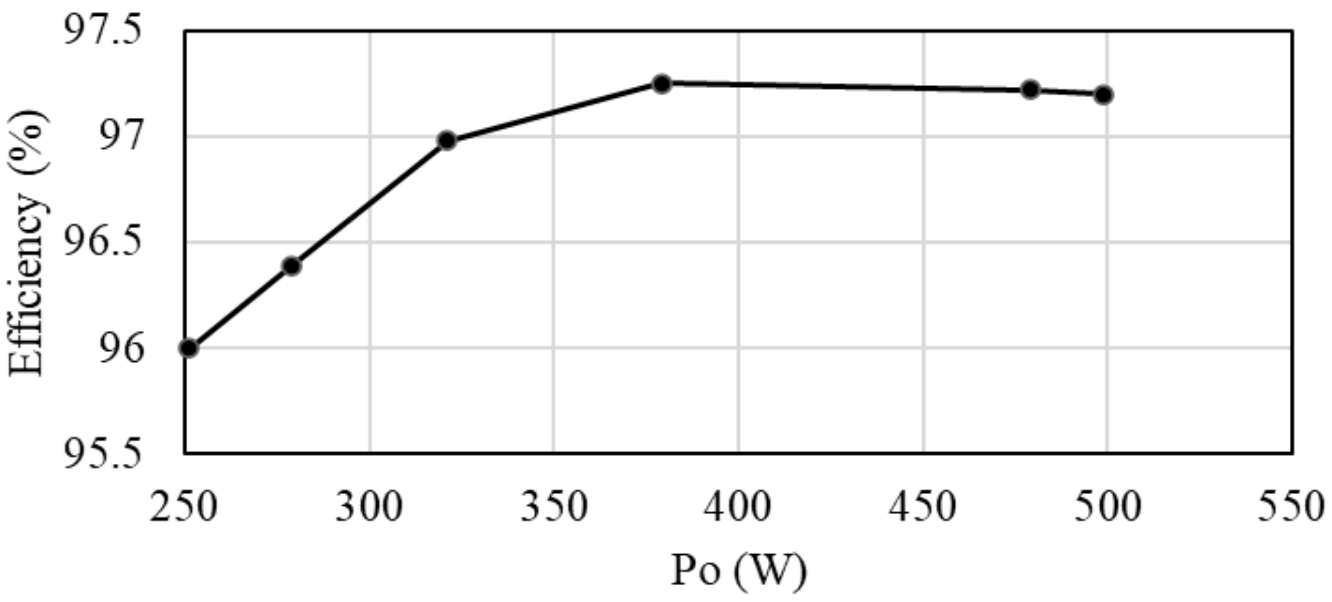

Fig. 17. Measured efficiency curve for the IZCC operation using the designed IM.

Fig. 17 shows the measured efficiency curve of the IZCC converter with the designed IM for different load powers ($P_O$). This figure shows that, as load power increases, efficiency rises and levels off at higher powers. At low powers, turn-off switching and core losses are considerable relative to the transferred power, and as power increases, conduction losses rise, keeping the efficiency curve almost constant at high power. The converter's peak efficiency is 97.25 %.

Table VIII shows a comparison between the proposed IM in this paper and the other discrete magnetic approaches. As shown, at the same switching frequency, the IZCC can achieve higher peak efficiency than the discrete design, and the IM volume per transferred power is lower than in the discrete approach. Therefore, utilizing the IM approach has the benefits of saving volume and improving efficiency.

TABLE VIII. A COMPARISON BETWEEN THE PROPOSED IM AND THE OTHER DISCRETE MAGNETICS IN IZCC

| Magnetic type | Frequency (*k*Hz) | Core volume per power ($mm^3$/W) | Peak efficiency % |
|---|---|---|---|
| Proposed IM | 150 | 104.29 | 97.25 |
| Discrete [33] | 150 | 185.19 | 97.1 |
| Discrete [31] | 40 | 261.86 | 97.8 |

## VII. CONCLUSIONS

This article proposes a new integrated magnetics design comprising six magnetic windings, including CI, RIs, and a transformer. The IM design was such that both DC and AC flux cancellation could be achieved, reducing the core volume and losses. In this regard, a U-core structure was proposed, which suggests that the DC flux can easily be cancelled, and the AC flux is cancelled in a limited range, such that the current ripple in the CI winding is large enough for achieving ZVS on the switches, as the CI current ripples are effective in ZVS achievement. It needs to be considered when the converter operates at a high switching frequency. Therefore, special attention needs to be given to winding arrangements, CC values, and flux directions. Based on these considerations, the IM was built and tested with the IZCC hardware, demonstrating that the IM design concept works and that ZVS can be achieved across all switches, resulting in a peak efficiency of 97.25% for the IZCC converter with IM. The hardware results show a 43% reduction in core volume and an efficiency improvement when using the IM design compared to the discrete approach.